\documentclass[aps,twocolumn,amsmath,amssymb,showpacs,fixfloats,
superscriptaddress,unsortedaddress]{revtex4}

\usepackage{graphicx}

\begin{document}

\title{Optical Sum Rule in Strongly Correlated Systems}
\author{E. Z. Kuchinskii, N. A. Kuleeva, I. A. Nekrasov, M. V. Sadovskii
\footnote{E-mail: sadovski@iep.uran.ru}}
\affiliation{
Institute for Electrophysics, Russian Academy of Sciences, Ural Branch,
Ekaterinburg 620016, Russia}

\begin{abstract}

We discuss the problem of a possible ``violation'' of the optical sum
rule in the normal (non superconducting) state of strongly correlated electronic 
systems, using our recently proposed DMFT+$\Sigma$ approach, applied to two 
typical models: the ``hot -- spot'' model of the pseudogap state and disordered 
Anderson -- Hubbard model.  We explicitly demonstrate that the general Kubo 
single band sum rule is satisfied for both models. However, the optical 
integral itself is in general dependent on temperature and characteristic 
parameters, such as pseudogap width, correlation strength and disorder 
scattering, leading to effective ``violation'' of the optical sum rule, which 
may be observed in the experiments.

\end{abstract}

\maketitle

\section{Introduction}

Many years ago Kubo \cite{Kubo} has proven the general sum rule for diagonal 
dynamic (frequency dependent) conductivity $\sigma(\omega)$, which holds 
for any system of charged particles irrespective of interactions, temperature 
or statistics. This sum rule is usually written as:  
\begin{equation} 
\frac{2}{\pi}\int_{0}^{\infty}Re\sigma({\omega})d\omega 
=\sum_r\frac{n_re^2_r}{m_r} 
\label{Kub_sum} 
\end{equation} 
where $r$ specifies the type of charged particles, $n_r$ and $e_r$ are the 
respective densities and charges.

For the system of electrons in a solid Eq. (\ref{Kub_sum}) takes the form:
\begin{equation}
\int_{0}^{\infty}Re \sigma({\omega}) d\omega =
\frac{\omega_{pl}^2}{8}
\label{f_sum_el}
\end{equation}
where $n$ is the density of electrons and $\omega_{pl}^2=\frac{4\pi ne^2}{m}$
is the plasma frequency.

However, in any real experiment we are not dealing with an infinite range of
frequencies. If one considers electrons in a crystal and limits himself to 
the electrons in a particular (e.g. conduction) band, neglecting interband
transitions, the general sum rule (\ref{f_sum_el}) reduces to the single band
sum rule of Kubo \cite{Kubo, vdM}:
\begin{equation}
W = \int_0^{\omega_c} Re \sigma(\omega) d\omega =
f(\omega_c) \frac{\pi e^2}{2}\sum_{\bf p}
\frac{\partial^2\varepsilon_p}{\partial p_x^2}n_p 
\label{1}
\end{equation}
where $\varepsilon_p$ is the bare dispersion as defined by the 
effective single band Hamiltonian, while $n_p$ is the momentum distribution 
function (occupation number), which is in general defined by the 
{\em interacting} retarded electronic Green's function $G^R(\varepsilon,{\bf  p})$ 
\cite{Pepin,Toshi}:  
\begin{equation} 
n_p=-\frac{1}{\pi}\int_{-\infty}^{\infty} d\varepsilon 
n(\varepsilon)Im G^R(\varepsilon,{\bf p})
\label{dist_fun}
\end{equation}
where $n(\varepsilon)$ is the usual Fermi distribution.
In Eq. (\ref{1}) $\omega_c$ represents an ultraviolet cut-off,  
frequency, which is assumed to be larger than the bandwidth 
of the low energy band, but smaller than the gap to other bands.
The function $f(\omega_c)$ accounts for the cut-off 
dependence, which arises from the presence of Drude spectral weight beyond 
$\omega_c$ \cite{Maksimov} and is unity if we formally set $\omega_c$ to 
infinity while ignoring the interband transitions.

Although the general sum rule is certainly preserved, the optical integral
$W(\omega_c, T)$ is not a conserved quantity since both $f(\omega_c)$ \cite{Maksimov}
and $n_p$ \cite{Toshi,Mike00} depend on temperature $T$, 
and also on details of interactions \cite{Pepin}. This dependence of $W$ on $T$ and
other parameters of the system under study has been termed the 
``sum rule violation''. It was actively studied experimentally, especially
in cuprates, where pronounced anomalies were observed both in
c-axis and in-plane conductivity both in normal and superconducting 
states \cite{Basov,Molegraaf,Santander,Bi2223,Erik,RMP}.

The finite cut-off effects were extensively studied in several theoretical 
papers on the $T$ dependence of the optical integral \cite{Toshi,Maksimov,Maks}. In Refs. \cite{Maksimov,Maks}, the 
effect of the cut-off was considered in the context of electrons coupled to 
phonons. In a simple Drude model, $\sigma(\omega) = 
(\omega_{pl}^2/4\pi)/(1/\tau-i\omega)$ and the sum rule can only be ``violated''
due to the presence of $f(\omega_c)$. Integrating over $\omega$ and expanding 
for  $\omega_c \tau >> 1$, one can see that 
\begin{equation}
f(\omega_c) = \left(1-\frac{2}{\pi}\frac{1}{\omega_c \tau}\right)
\label{max}
\end{equation}
For infinite cut-off, $f(\omega_c) =1$ and  $W = \omega_{pl}^2/8$, 
but for a finite cut-off $f(\omega_c)$ contains the term proportional
to $1/\omega_c \tau$.  If $1/\tau$ changes with $T$, then one obtains 
a sum rule ``violation'' even if $\omega_{pl}$ is $T$ independent
\cite{Maksimov,Maks}. Other aspects of cut-off dependence were discussed 
recently in detail in Ref. {\cite{vdM}}.

In the present study we neglect the cut-off effects in optical integral from
the outset. Our goal is to study $W$ dependence on 
$T$ and a number of interaction parameters, determining the
electronic properties of strongly correlated systems, such as cuprates. In 
this context we shall discuss the problem of a possible ``violation'' of the 
optical sum rule in the normal (non superconducting) state of strongly 
correlated electronic systems, using our recently proposed DMFT+$\Sigma$ 
approach \cite{jtl,cm05,FNT}, as applied to dynamic conductivity in two typical 
models of such systems: the ``hot -- spot'' model of the pseudogap 
state \cite{SkOpt} and disordered Anderson -- Hubbard model \cite{HubDis}. 
Our aim is both to check the consistency 
of DMFT+$\Sigma$ approach as applied to calculations of optical conductivity,
as well as to demonstrate rather important dependences of the optical
integral $W$ not only on $T$, but also on such important characteristics as
pseudogap width, disorder and correlation strength, making (single band) sum 
rule ``violation'' rather ubiquitous for any strongly correlated system, even
neglecting the cut-off effects.  

\section{Optical sum rule in the generalized DMFT+$\Sigma$ approach}

Characteristic feature of the general sum rule as expressed by Eqs. (\ref{1}),
(\ref{dist_fun}) is that the integral $W$ over frequency in the l.h.s. is
calculated via two-particle property (dynamic conductivity, determined by
two-particle Green's function, in general, with appropriate vertex corrections), 
while the r.h.s. is determined by the single-particle characteristics, such as 
bare dispersion and occupation number (\ref{dist_fun}) (determined by a
single-particle Green's function). Thus, checking the validity of this sum 
rule, we are in fact thoroughly checking the consistency of any theoretical 
approach, used in our model calculations.

Our generalized dynamical mean field theory  (DMFT+$\Sigma$) 
approach \cite{jtl,cm05,FNT}, supplying the standard dynamical mean field
theory (DMFT) \cite{pruschke,georges} with an additional
``external'' self-energy $\Sigma$ (due to any kind of interaction outside the
scope of DMFT, which is exact only in infinite dimensions), provides an effective
method to calculate both single- and two-particle properties 
\cite{SkOpt,HubDis}. The consistency check of this new approach is obviously
of great interest by itself. We shall also see, that it gives a kind of a new
insight in the problem of sum-rule ``violation''.

\subsection{Pseudogap state, the ``hot spots'' model}

Pseudogap phenomena in strongly correlated systems have essential spatial
length scale dependence \cite{MS}. To merge pseudogap physics and strong 
electron correlations we have generalized the dynamical-mean field theory 
\cite{pruschke,georges} by inclusion of the dependence
on correlation length of pseudogap fluctuations via additional (momentum
dependent) self-energy $\Sigma_{\bf p}(\varepsilon)$. 
This self-energy $\Sigma_{\bf p}(\varepsilon)$ 
describes non-local dynamical correlations induced either by short-ranged 
collective SDW-like antiferromagnetic spin or CDW-like charge fluctuations 
\cite{Sch,KS}. 

To calculate $\Sigma_{\bf p}(\varepsilon)$ in two-dimensional ``hot spots'' 
model \cite{MS} for an electron moving in the random field of pseudogap 
fluctuations (considered to be static and Gaussian) with dominant
scattering momentum transfers of the order of characteristic
vector ${\bf Q}=(\pi/a,\pi/a)$  ($a$ is the lattice spacing), 
we used \cite{cm05,FNT} the recursion procedure proposed in Refs.~\cite{Sch,KS}, which is
controlled by two main physical characteristics of the pseudogap state:
$\Delta$ (pseudogap amplitude), which characterizes the energy scale of the 
pseudogap, and $\kappa=\xi^{-1}$ -- the inverse correlation length of short range
SDW (CDW) fluctuations. Both parameters $\Delta$ and $\xi$, determining 
pseudogap behavior, can in principle be calculated from the microscopic model 
at hand \cite{cm05}. 

Weakly doped one-band Hubbard model with repulsive Coulomb interaction $U$ on a 
square lattice with nearest and next nearest neighbour hopping was numerically
investigated within this generalized DMFT+$\Sigma$ self-consistent approach,
as described in detail in Refs. \cite{jtl,cm05,FNT}.  

Briefly, the DMFT+$\Sigma$ self-consistent loop looks like as follows.
First we guess some initial local (DMFT) electron self-energy $\Sigma(\varepsilon)$.
Second we compute  the ${\bf p}$-dependent ``external''
self-energy $\Sigma_{\bf p}(\varepsilon)$ which is in general case
a functional of $\Sigma(\varepsilon)$. Then neglecting interference effects between 
the self-energies (which in fact is the major assumption of our approach)
we can set up and solve the lattice problem of DMFT \cite{pruschke,georges}.
Finally we define effective Anderson single impurity problem which 
is to be solved by any ``impurity solver'' (we mostly use numerical
renormalization group - NRG) to close DMFT+$\Sigma$ equations. 

The additive form of self-energy is in fact an advantage of 
our approach \cite{jtl,cm05,FNT}.  It allows one to preserve the set of 
self-consistent equations of standard DMFT \cite{pruschke,georges}.
However there are two distinctions from conventional DMFT.
During each DMFT iteration we recalculate corresponding  {\bf {p}}-dependent 
self-energy $\Sigma_{\bf p}(\mu,\varepsilon,[\Sigma(\omega)])$ via an
approximate scheme, 
taking into account interactions with collective 
modes or order parameter fluctuations, and the local Green function 
$G_{ii}(i\omega)$ is ``dressed'' by $\Sigma_{\bf p}(\varepsilon)$ at each step. 
When input 
and output Green's functions (or self-energies) converge to each other (with 
prescribed accuracy) we consider the obtained solution to be selfconsistent.  
Physically it corresponds to the account of some ``external'' (e.g. pseudogap) 
fluctuations, characterized by an important length scale $\xi$,
into fermionic ``bath'' surrounding the effective Anderson impurity of the 
usual DMFT.
Both cases of strongly correlated metals and doped Mott insulators
were considered \cite{cm05,FNT}. 
Energy dispersions, quasiparticle damping, spectral functions and ARPES 
spectra calculated within DMFT+$\Sigma$, all show a pseudogap effects close 
to the Fermi level of quasiparticle band.

In Ref. \cite{SkOpt} this DMFT+$\Sigma$ procedure was generalized to 
calculate two-particle properties, such as dynamic conductivity, using
previously developed recursion procedure for vertex corrections due to
pseudogap fluctuations \cite{SS02}, producing typical pseudogap anomalies
of optical conductivity and dependence of these anomalies on correlation strength
$U$. Below we use the approach of Ref. \cite{SkOpt} to investigate the sum-rule
in ``hot spots'' model. 

To calculate optical integral $W$ we have just used the
conductivity data of Ref. \cite{SkOpt} (extended to a wider 
frequency range needed to calculate $W$), while the r.h.s of (\ref{1}) was 
recalculated, using recursion relations for $\Sigma_{\bf p}(\varepsilon)$ and 
the whole self-consistency DMFT+$\Sigma$ loop. 
All calculations have been done for a tight-binding ``bare'' spectrum
on the square lattice, with the nearest neighbor transfer integral $t$ and
next nearest neighbor transfer integral $t'$.

\begin{figure}
\includegraphics[width=\columnwidth] {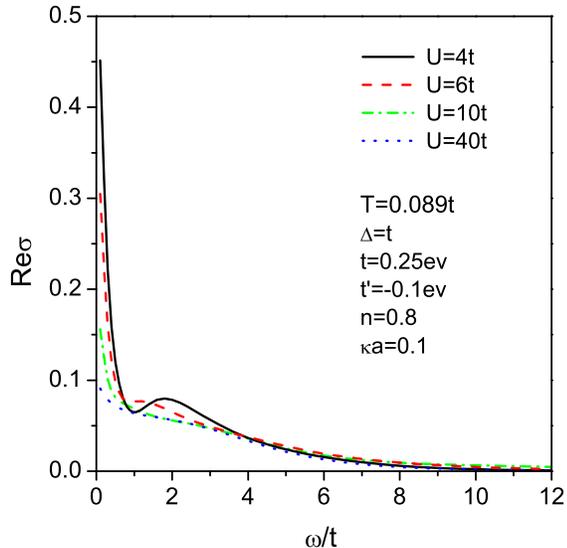}
\caption{Real part of optical conductivity for strongly 
correlated system in the pseudogap state ($t'=-0.4t$, $t=0.25$ eV, $T=0.089t$) 
in DMFT+$\Sigma_{\bf p}$ approximation --- $U$ dependence. Band filling
$n=0.8$, pseudogap amplitude $\Delta=t$,  correlation length  
$\xi=10 a$. Conductivity is given in units of $\sigma_0=\frac{e^2}{\hbar}$.
}
\label{fig1}
\end{figure}

In Fig. \ref{fig1} we present our typical data for the real part of 
conductivity ($t'=-0.4t$, $t=0.25$ eV, band filling $n=0.8$, temperature 
$T=0.089t$) for different values of Hubbard interaction $U=4t,\ 6t,\ 10t,\ 
40t$ and fixed pseudogap amplitude $\Delta=t$ (correlation length
$\xi=10 a$).  It is obvious from these data, that optical integral $W$ is 
different for all of these curves, actually its value drops with the growth 
of $U$ (along with damping of pseudogap anomalies \cite{SkOpt}). 
However, the single band optical sum-rule (\ref{1}) is satisfied within our 
numerical accuracy, as seen from Table I. The small ``deficiency'' in the
values of $W$ in Table I is naturally due to a finite frequency 
integration interval over conductivity data of Fig. \ref{fig1}.

\begin{table}
\label{Tab1}
\caption {Single-band optical sum rule check in the ``hot-spots'' model. 
$U$ - dependence.
Optical integral in units of $\frac{e^2}{\hbar}t$.}
\begin{tabular}{| c | c | c |}
\hline
$U$    &  $\frac{\pi e^2}{2}\sum_{\bf p}\frac{\partial^2\varepsilon_p}{\partial p_x^2}n_p$  & $W = \int_0^{\infty} Re \sigma(\omega) d\omega $ \\
\hline
$U=4t$  & 0.456 & 0.408 \\
\hline
$U=6t$  & 0.419 & 0.387 \\
\hline
$U=10t$ & 0.371 & 0.359  \\
\hline
$U=40t$ & 0.323 & 0.306 \\
\hline
\end{tabular}
\end{table}

\begin{figure}
\includegraphics[clip=true,width=\columnwidth]{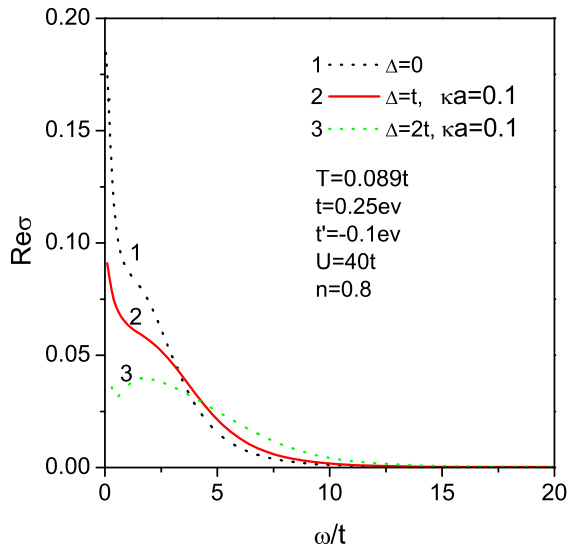}
\caption{Real part of optical conductivity for doped Mott 
insulator ($U=40t$, $t'=-0.4t$, $t=0.25$ eV, $T=0.089t$) 
in DMFT+$\Sigma_{\bf p}$ approximation for different values of 
pseudogap amplitude $\Delta=0$, $\Delta=t$, $\Delta=2t$. 
Correlation length $\xi=10 a$, band filling factor $n=0.8$.
} 
\label{fig2} 
\end{figure}

In Fig. \ref{fig2} we show the real part of optical conductivity for doped Mott 
insulator (fixed $U=40t$, $t'=-0.4t$, $t=0.25$ eV, band filling $n=0.8$, 
$T=0.089t$) for different values of 
pseudogap amplitude $\Delta=0$, $\Delta=t$, $\Delta=2t$. 
Correlation length is again $\xi=10 a$ and band filling factor $n=0.8$.
The ``violation'' of sum-rule here is especially striking ---
optical integral obviously drops with the growth of $\Delta$. However, again
the single band optical sum-rule (\ref{1}) is strictly valid, as seen from
Table II.

\begin{table}
\label{Tab2}
\caption {Single-band optical sum rule check in the ``hot spots'' model. 
$\Delta$ - dependence.
Optical integral in units of $\frac{e^2}{\hbar}t$.}
\begin{tabular}{| c | c | c |}
\hline
$\Delta$    &  $\frac{\pi e^2}{2}\sum_{\bf p}\frac{\partial^2\varepsilon_p}{\partial p_x^2}n_p$  & $W = \int_0^{\infty} Re \sigma(\omega) d\omega $ \\ 
\hline
$\Delta=0$ & 0.366 & 0.36  \\
\hline
$\Delta=t$ & 0.314 & 0.304  \\
\hline 
$\Delta=2t$ & 0.264 & 0.252 \\  
\hline
\end{tabular}
\end{table}

\begin{figure}
\includegraphics[clip=true,width=\columnwidth]{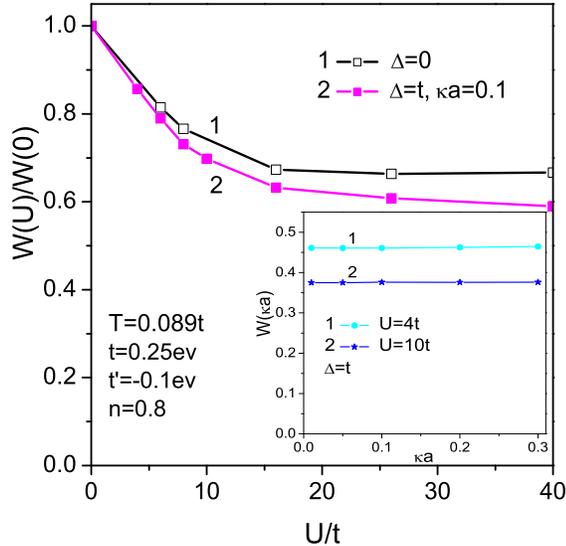}
\caption{Dependence of normalized optical integral on
correlation strength $U$ in the pseudogap state. All other parameters
are listed in the figure. At the insert --- correlation length dependence
of optical integral in units of $\frac{e^2}{\hbar}t$ .} 
\label{fig3} 
\end{figure}

\begin{figure}
\includegraphics[clip=true,width=\columnwidth]{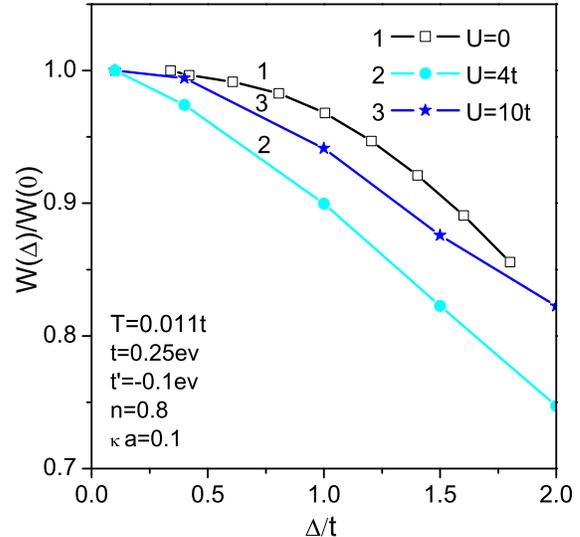}
\caption{Dependence of the normalized optical integral on the
pseudogap amplitude $\Delta$. The other parameters are listed in the figure.}  
\label{fig4} 
\end{figure}

\begin{figure}
\includegraphics[clip=true,width=\columnwidth]{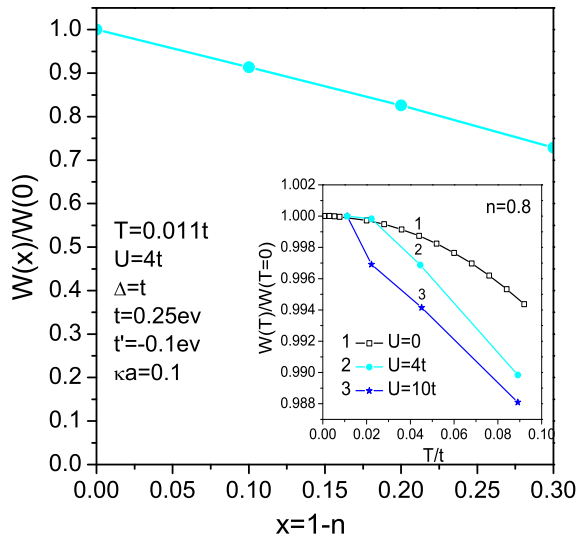}
\caption{Dependence of the normalized optical integral on
hole doping in the pseudogap state. At the insert --- temperature dependence.
All other parameters are listed in the figure}  
\label{fig5} 
\end{figure}

To study the details of sum-rule ``violation'', i.e. the dependence of the
optical integral $W$ on the parameters of the model, we performed extensive
calculations of the appropriate dependences of the r.h.s. of Eq. (\ref{1})
and optical integral $W$ on the temperature $T$, doping, pseudogap amplitude 
$\Delta$, correlation length of pseudogap fluctuations $\xi=\kappa^{-1}$ and 
correlation strength $U$. Some of the results are presented in Figs. 
\ref{fig3} -- \ref{fig5}.

Typical dependence of the (normalized)
optical integral on correlation strength $U$ is shown in  Fig. \ref{fig3} for
two values of $\Delta$. 
We can see rather significant drop of $W$ with the growth of $U$.
As to correlation length dependence, which is shown at the insert Fig. \ref{fig3}, 
it was found to be very weak 
(practically negligible) in a whole region of realistic values of $\xi$,
so that we shall not discuss it further. 
Dependence of $W$ on the pseudogap amplitude $\Delta$ (for several 
values of $U$) is shown in Fig. \ref{fig4}. 
Typical doping dependence, which reflects just the dependence of 
the square of the plasma frequency $\omega_{pl}^2$ on doping, is given in Fig. \ref{fig5}. 
In all cases other, the change of the relevant parameters of the model lead to rather
significant drop in the values of $W$. 
As to the temperature dependence (shown in the insert on Fig. \ref{fig5}) it
is rather weak, quadratic in $T$ and quite similar to that found in 
Refs. \cite{Toshi}.

Basically these results show, that the value of the optical integral depends
on all the major parameters of the model and, in this sense, its value is
not universal, so that the optical sum rule is significantly ``violated'',
if we restrict ourselves with a single-band contribution.

\subsection{Disordered Anderson -- Hubbard model}

In Ref. \cite{HubDis} we have applied DMFT+$\Sigma$ approximation
to calculate the density of states, optical conductivity and phase 
diagram of strongly correlated and strongly disordered paramagnetic 
Anderson--Hubbard model, with Gaussian site disorder.  
Strong correlations were accounted by DMFT, while 
disorder was taken into account via the appropriate generalization of 
self-consistent theory of localization \cite{VW,WV,MS83,MS86}.  
We considered the three-dimensional system with semi-elliptic density 
of states.  
Correlated metal, Mott insulator and correlated Anderson insulator phases 
were identified via the evolution of density of states and dynamic 
conductivity, demonstrating both Mott-Hubbard and Anderson metal-insulator 
transitions and allowing the construction of complete zero-temperature phase 
diagram of Anderson--Hubbard model. 

For ``external'' self-energy entering DMFT+$\Sigma$ loop we have used 
the simplest possible approximation (neglecting ``crossing'' diagrams for 
disorder scattering), i.e. just the self--consistent Born approximation, 
which in the case of Gaussian site energies disorder takes the usual 
form:
\begin{equation}
\Sigma(\varepsilon)=\Delta^2\sum_{\bf p}G(\varepsilon,{\bf p})
\label{BornSigma}
\end{equation}
where $\Delta$ denotes now the amplitude of site disorder.
 
Calculations of optical conductivity are considerably simplified \cite{HubDis}
due to the fact, that there are no contributions to conductivity due to vertex 
corrections, determined by local Hubbard interaction. Finally, conductivity is 
essentially determined by the generalized diffusion coefficient, 
which is obtained from the appropriate generalization of self-consistent 
equation of Refs. \cite{VW,WV,MS83,MS86}, which is to be solved in conjunction 
with DMFT+$\Sigma$ loop. 

\begin{figure}
\includegraphics[clip=true,width=\columnwidth]{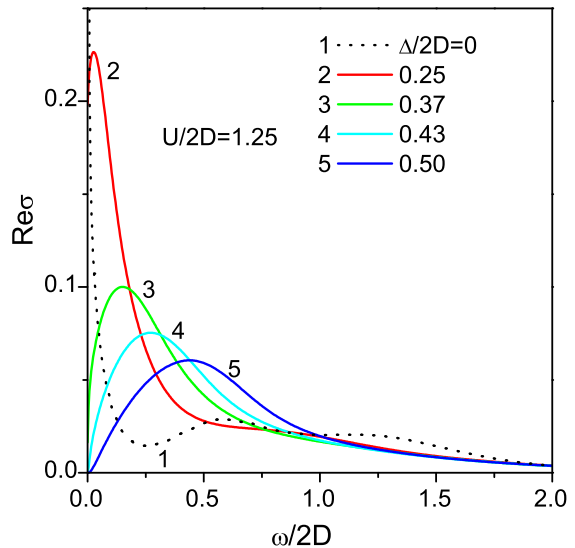}
\caption{Real part of dynamic conductivity for half--filled 
Anderson--Hubbard model for different degrees of disorder $\Delta$, 
and $U=2.5D$, typical for correlated metal. Lines 1,2 are for metallic
phase, line 3 corresponds to the mobility edge (Anderson transition), 
lines 4,5 correspond to correlated Anderson insulator.
Conductivity is in units of $\frac{e^2}{\hbar a}$.}  
\label{fig6} 
\end{figure}

In Fig. \ref{fig6} we show typical results for the real part of dynamic 
conductivity of a correlated metal described by the half--filled 
Anderson--Hubbard model (with bandwidth $2D$) for different degrees of 
disorder $\Delta$, and $U=2.5D$, and demonstrating continuous transition  
to correlated Anderson insulator with the growth of disorder.

Here again the direct check shows that the single band optical sum-rule 
(\ref{1}) is obeyed within our numerical accuracy, as seen from Table III.
At the same time, optical integral $W$ itself obviously changes with
disorder.

\begin{table}
\label{Tab3}
\caption {Single-band optical sum rule check in Anderson -- Hubbard model. 
$\Delta$ - dependence.
Optical integral in units of $\frac{2e^2}{\hbar a}D$.}
\begin{tabular}{| c | c | c |}
\hline
$\Delta/2D$    &  
$\frac{\pi e^2}{2}\sum_{\bf p}\frac{\partial^2\varepsilon_p}{\partial p_x^2}n_p$  & $W = \int_0^{\infty} Re \sigma(\omega) d\omega $ \\ 
\hline 
$0$  & 0.063 & 0.064 \\ 
\hline 
$0.25$  & 0.068 & 0.07 \\ 
\hline
$0.37$ & 0.06 & 0.056  \\
\hline
$0.5$ & 0.049  & 0.05 \\
\hline
\end{tabular}
\end{table}

\begin{figure}
\includegraphics[clip=true,width=\columnwidth]{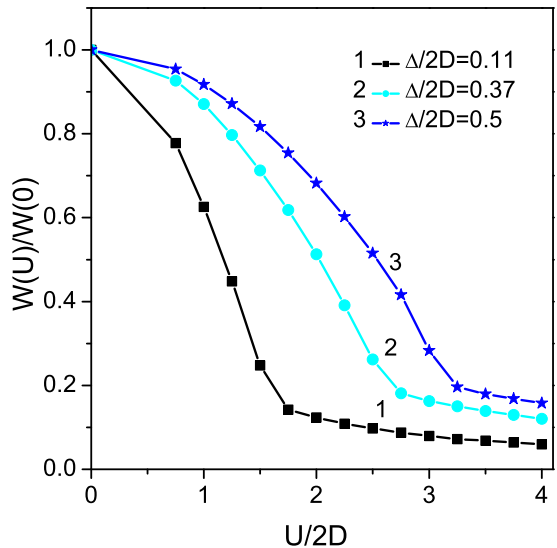}
\caption{Dependence of the normalized optical 
integral on correlation strength in Anderson-Hubbard model for different
degrees of disorder $\Delta$ (1,2 -- strongly disordered metal, 3 -- 
correlated Anderson insulator). } 
\label{fig7} 
\end{figure}

\begin{figure}
\includegraphics[clip=true,width=\columnwidth]{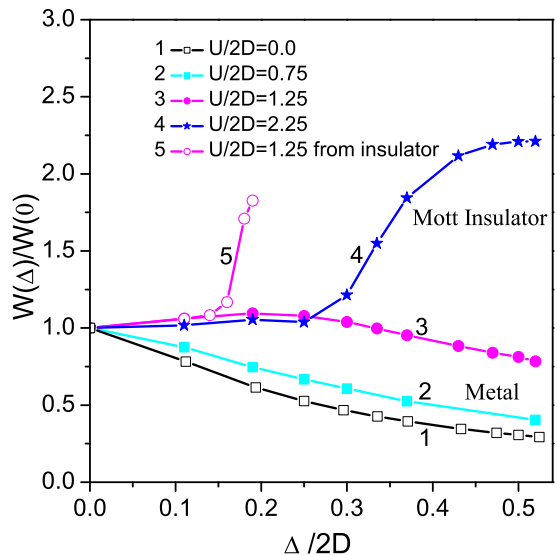}
\caption{Disorder dependence of the normalized optical
integral in Anderson-Hubbard model for different values of Hubbard
interaction $U$. Lines 1,2,3 -- correlated metal, transforming into
Anderson insulator. Line 4 -- Mott insulator state obtained with the growth 
of $U$ from correlated metal, line 5 -- Mott insulator obtained with
diminishing $U$ in hysteresis region of the phase diagram.} 
\label{fig8} 
\end{figure}

\begin{figure}
\includegraphics[clip=true,width=\columnwidth]{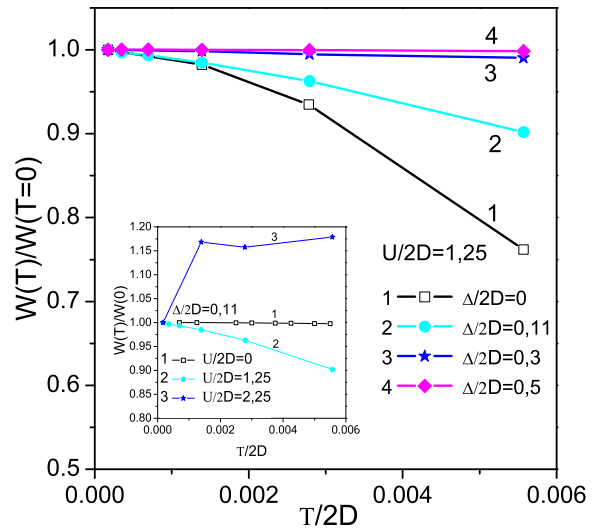}
\caption{Temperature dependence of normalized optical
integral in Anderson-Hubbard model for different degrees of disorder.  
At the insert -- similar dependence at fixed disorder, but for different
values of Hubbard interaction $U$, line 3 here corresponds to disordered
Mott insulator.} 
\label{fig9} 
\end{figure}

Again, to study the details of this sum-rule ``violation'', i.e. the 
dependence of $W$ on the parameters of Anderson-Hubbard model, we 
performed detailed calculations of its dependences on the temperature $T$, 
disorder amplitude $\Delta$ and correlation strength $U$. Some of the 
results are presented in Figs.  \ref{fig7} -- \ref{fig9}.

In Fig. \ref{fig7} we show the dependence of normalized optical integral on
Hubbard $U$, for different degrees of disorder (both for strongly 
disordered metal and correlated Anderson insulator). It is seen that in all 
cases the growth of correlation strength leads to rather sharp drop of $W$ 
in metallic state, which becames mush slower in Mott insulator. 

In Fig. \ref{fig8} we present similar dependences on disorder strength $\Delta$.
In metallic state optical integral generally drops with the growth of 
disorder, while an opposite behavior is observed if we start from Mott insulator (both
obtained with the growth of $U$ from metallic state and under diminishing $U$ 
in hysteresis region of the phase diagram \cite{HubDis}). Note the absence of
any significant changes in the immediate vicinity of critical disorder 
$\Delta_c/2D=0.37$, corresponding to Anderson metal -- insulator transition.
At the same time it should be noted that the most significant growth of the optical
integral takes place as the system transforms into disorder induced metallic 
 state, obtained from Mott insulator,  as observed in Ref.\cite{HubDis}.

In Fig. \ref{fig9} we show the temperature dependence of the normalized optical 
integral, for different degrees of disorder. 
In Anderson -- Hubbard model it appears to be significantly
stronger, than in ``hot spots'' model
(see above), and decreases with the growth of disorder. Moreover, while in
relatively weakly correlated state it is qualitatively the same -- optical
integral diminishes with the growth of $T$, it actually grows in 
disordered Mott insulator, as seen from line 3 at the insert in Fig. 9. 

Again, as in the case of the pseudogap ``hot spots'' model, these results for
Anderson -- Hubbard model clearly demonstrate the value of the optical integral 
is not universal and depends on all the major 
parameters of the model and, so the single band optical sum rule is 
strongly ``violated''.

\section{Conclusion}

Based on DMFT+$\Sigma$ approach, we have studied the single band optical sum
rule for two typical strongly correlated systems, which are outside the scope
of the standard DMFT: (i) the ``hot spots'' model of the pseudogap state,
which takes into account important nonlocal correlations due to AFM(CDW)
short-range order fluctuations and (ii) Anderson-Hubbard model, which 
includes strong disorder effects, leading to disorder induced metal-insulator
(Anderson) transition, alongside with Mott transition. 

We have explicitly demonstrated that the single band optical sum rule 
(\ref{1}) is satisfied for both models, confirming the self-consistency of
DMFT+$\Sigma$ approach for calculation of two-particle properties.

However, the optical integral $W = 2\int_0^{\infty} Re \sigma(\omega) d\omega $,
entering the single band sum rule (\ref{1}) is non universal and depends on
the parameters of the model under consideration. Most of the previous studies
addressed its (relatively weak) temperature dependence. Here we have 
analyzed dependences on essential parameters of our models, showing
that these may lead to rather strong ``violations'' of the optical sum rule.
As most of the parameters under discussion may be varied in different kinds
of experiments, these dependences should be taken into account in the analysis 
of optical experiments on strongly correlated systems.

\section{Acknowledgements}

We thank Thomas Pruschke for providing us the NRG code.
This work is supported by RFBR grants 08-02-00021, 08-02-00712, 
RAS programs ``Quantum macrophysics'' and ``Strongly correlated electrons in 
semiconductors, metals, superconductors and magnetic materials''. 
IN is also supported the Grant of the President of Russia MK.2242.2007.2 and 
Russian Science Support Foundation.

\end{document}